\documentclass[9pt,twocolumn,twoside]{opticajnl}
\journal{opticajournal} 

\setboolean{shortarticle}{true}

\usepackage{lineno}
\usepackage{upgreek}

\title{Laser fractional frequency instability at $\mathbf{4\times 10^{-17}}$ with a room temperature optical reference cavity}

\author[1]{Adam L. Parke}
\author[1]{Eve Clulow}
\author[1]{Wei Huang}
\author[1]{Namneet Kaur}
\author[1]{Reinhard Karembera}
\author[1]{Jacques-Olivier Gaudron}
\author[1]{Xi Zhang}
\author[1]{Matias Risaro}
\author[1]{Jacob Tunesi}
\author[1]{Henry Bourne}
\author[1,*]{Marco Schioppo}

\affil[1]{National Physical Laboratory (NPL), Teddington, TW11 0LW, UK}

\affil[*]{marco.schioppo@npl.co.uk}

\begin{abstract}
Ultrastable lasers play a key role in optical frequency metrology, setting measurement speed and ultimately impacting both stability and accuracy of optical frequency standards. Here we demonstrate laser fractional frequency instability at $\mathbf{{4}\times10^{-17}}$ and laser frequency linewidth of $\mathbf{12}\,$mHz full width at half maximum, employing a 68~cm long optical reference cavity operating at room temperature. To the best of our knowledge, both frequency instability and linewidth are the lowest ever reported for a room temperature system. This work highlights that state-of-the-art frequency stability and spectral purity are achievable at room temperature, making them accessible to a broader range of users.
\end{abstract}

\setboolean{displaycopyright}{false} 

\begin{document}

\maketitle
The advent of low frequency-noise lasers with linewidths approaching those of narrow atomic transitions in the optical domain has been transformational and has contributed to opening the field of optical frequency metrology~\cite{Ludlow:15}. Today's most stable lasers employ optical Fabry-Pérot cavities as phase and frequency references, and use the Pound-Drever-Hall (PDH) stabilization method~\cite{Pound:46, Drever:83} to transfer the cavities' fractional length stability into optical fractional frequency stability. Over time, the design of these optical cavities has evolved to enhance their length stability by minimizing the sensitivity to environmental sources of noise, such as temperature variations and acceleration noise~\cite{Nazarova:06, Chen:06}. If the effect of all external sources of noise is made negligible, the cavity length stability reaches a limit set by the Brownian thermal noise introduced by the high reflectivity coating of the mirrors, with the following scaling law~\cite{Numata:04, Jiang:11, KesslerTherm:12}
\begin{equation}
\sigma_{y}\propto\frac{1}{L_{\text{cav}} D_{\text{beam}}}\sqrt{\frac{k_{\text{B}}T}{E\,Q}}\,, 
\label{eq:1}
\end{equation}
where $L_{\text{cav}}$ is the cavity spacer length, $D_{\text{beam}}$ is the ($e^{-2}$) diameter of the optical beam on the coating, $k_{\text{B}}$ is the Boltzmann constant, $T$ and $E$ the temperature (K) and the Young's modulus of the mirror substrate, and $Q$ the coating mechanical quality factor. Recent efforts in the research community have mostly been focused on the development of optical reference cavities operating at cryogenic temperatures ($\leq130\,$K)~\cite{Wiens:14, Matei:17, Kedar:23, Valencia:24, Chen:25, Lee:2026} to reduce the thermal noise, as suggested by Eq.~(\ref{eq:1}), with the additional benefit of using materials with a high Young's modulus and a low coefficient of thermal expansion (CTE) at these temperatures, such as silicon and sapphire. A high Young's modulus also reduces the sensitivity to acceleration noise. New types of coatings with higher mechanical quality factor have also been used~\cite{Kedar:23, Lee:2026}. However, operation at cryogenic temperature has drawbacks, such as the additional complexity, cost, equipment footprint, acoustic and acceleration noise introduced by the cryocooler, and limitations on continuous operation due to coolant refilling (in liquid or gas form) and regular mechanical servicing. \\ \indent
In this work we focus on the development of optical reference cavities operating at room temperature, which has historically been very successful~\cite{Ludlow:15,Young:99, Jiang:11, Hafner:15, Schioppo:17, Schioppo:22}, and we pursue it with the following three main motivations: to eliminate the previously mentioned drawbacks introduced by cryogenics; to realize a system capable of full continuous operation in order to support optical clocks in meeting the targets for the redefinition of the SI second~\cite{Dimarcq:24}; to use the ultrastable laser as a continuously-available optical flywheel to bridge optical clock data gaps towards an optical timescale \cite{Milner:19}. We target state-of-the-art stability and Eq.~(\ref{eq:1}) informs us that we need to leverage on long optical cavities with a large optical beam diameter on the mirrors.
\begin{figure*}[ht!]
\centering
\includegraphics[width=0.99\linewidth]{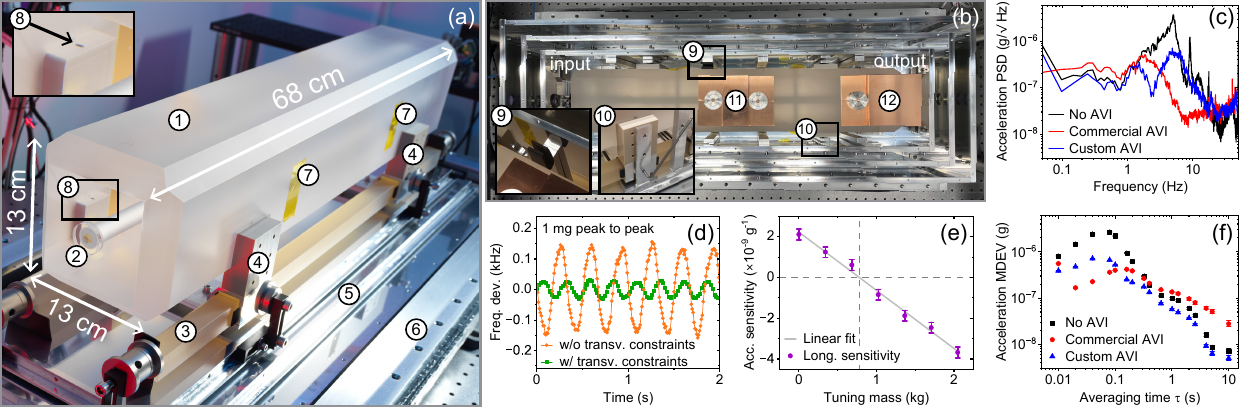}
\caption{(a) Cavity ULE68a at an earlier stage of assembly. The numbered parts are (1) ULE spacer, with dimensions 13~cm $\times$ 13~cm $\times$ 68~cm, and edges chamfered at $45^{\circ}$, (2) fused silica cavity mirror, (3) Zerodur bar, (4) PEEK supports, (5) base of thermal shields, (6) base of vacuum chamber, (7) venting holes, and (8) Viton disk, one on each PEEK support as its only point of contact with the cavity spacer. (b) Top view of ULE68a, with thermal shields mounted and acceleration insensitivity optimized. The numbered parts are (9) Viton spring pushing on the side of the cavity, (10) aluminum disk pushing on the cavity support. Copper and stainless steel tuning masses are placed on top of the cavity in the center (11) and at the output (12), to act on vertical and longitudinal acceleration sensitivity, respectively. Insets (9) and (10) show closer views of the transverse constraints. The Kapton tape over the venting holes in (7) and the Allan key in (10) were removed in the final configuration. (d) Frequency deviation caused by 1~mg peak-to-peak transverse excitation, with and without transverse constraints. (e) Longitudinal acceleration sensitivity measured with different tuning masses (12), showing a zero-crossing around 780~g. (c) and (f) are plots of longitudinal acceleration noise as power spectral density and modified Allan deviation respectively, with different active vibration isolation (AVI) configurations.}
\label{fig:1}
\end{figure*}
Driven by the above motivations, we developed a 68~cm long optical cavity, made in ultra-low-expansion (ULE) glass, with two optically-contacted mirrors of 10.2~m radius of curvature (ROC), resulting in a beam diameter at the mirrors of 1.9~mm. For brevity we will refer to this cavity as "ULE68a". We used a dielectric ion-beam-sputtering high-reflectivity coating, deposited on a mirror substrate of Suprasil 312 fused silica, which in combination with the spacer length and beam diameter gives a thermal noise limit for the instability calculated at $2.5\times10^{-17}$, using the noise model in~\cite{Numata:04, Jiang:11, KesslerTherm:12}. This coating type was chosen for its highly uniform reflectivity over a large beam area, high reproducibility and accessibility. As previously presented in~\cite{Parke:25}, this coating gives us a finesse of 410,000, which with a cavity length of 68~cm provides an optical storage time of about $300\,\mu\text{s}$ and a cavity linewidth of 540~Hz. This reduces the effect of residual amplitude modulation, which is further minimized far below the thermal noise by active stabilization. We stabilize the optical power to $20\,\mu$W before the cavity. We intra-cavity couple $40\,\%$ of the input power.
The design of ULE68a builds on the experience we acquired in the construction of two previous ULE systems with spacer lengths of 48.5~cm. The first system (referred to as "ULE48a"), whose performance has already been evaluated in~\cite{Schioppo:22}, has a 1~m ROC mirror and a flat mirror, yielding a thermal noise at $6\times10^{-17}$. The second system, "ULE48b", has two 10.2~m ROC mirrors with an estimated thermal noise floor at $5\times10^{-17}$. Both ULE48a and ULE48b have a cylindrical geometry, as suggested by finite-element-method (FEM) analysis to minimize the sensitivity to accelerations. In practice, their geometry has proven to be very challenging to realize in the glass machining stage. In this process, the glass blank is rotated along the optical axis and the excess material is lathed away. In our case, the rotation of a 48.5~cm long cavity with a weight of 8~kg produced mechanical vibrations that generated multiple defects during the machining, as ULE is prone to flaking under localized mechanical stress. Therefore this process could not be extended to fabrication of longer optical cavities. Additionally, defects increased the measured acceleration sensitivity with respect to FEM estimations.
\begin{figure*}[htb]
\centering
\includegraphics[width=0.94\linewidth]{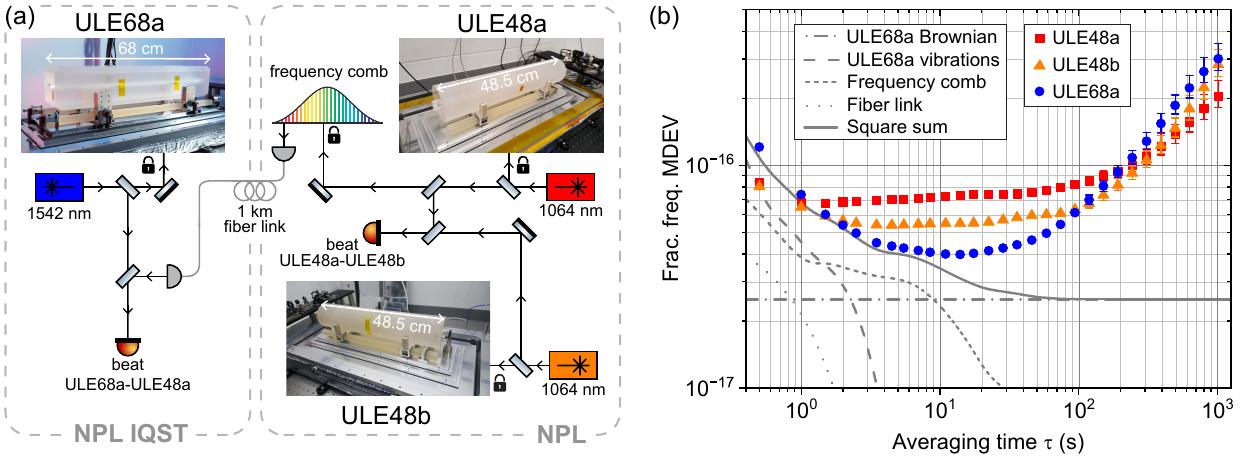}
\caption{(a) Schematic of the three-cornered hat (TCH) measurement set-up. ULE48a, ULE48b and the frequency comb are located in the same building, connected by short compensated fibers (not pictured). Beat \text{ULE48a–ULE48b} is produced from directly beating the 1064~nm lasers stabilized to the ULE48a and ULE48b cavities. The frequency comb is locked to ULE48a on the 1064~nm branch, and comb light at 1542~nm is sent to beat with light stabilized to ULE68a via a 1~km fiber link with path length compensation, to create the beat ULE68a–ULE48a. (b) Results of the TCH measurement of the three cavities over a 10~hour data set. Linear drift is removed for these stability evaluations. We display the main noise contributions for ULE68a, and their sum in quadrature.}
\label{fig:2}
\end{figure*}
In order to go beyond the above limitations, ULE68a has a cuboid geometry, as displayed in Fig.~\ref{fig:1}. This geometry allows for machining via milling rather than lathing, and for the cavity to be at rest during the machining process, held securely by its large flat surfaces. In addition, a milling tool has a larger contact area with the ULE glass than a lathe bit would have with the cylindrical geometry. These differences in the machining process ensure the mechanical forces are distributed over a larger area, and flaking of the glass is avoided. The resulting machined spacer shows no visible defects and a measured sensitivity to accelerations along the vertical axis in good agreement with FEM simulations. These predict that the vertical sensitivity critically depends on the distance between the horizontal plane intercepting the cavity optical axis and the plane defined by the supporting points~\cite{Nazarova:06, Chen:06}, thus strongly dependent on design and machining. The accurate machining enabled by this cuboid geometry has therefore been crucial to achieve a low vertical sensitivity measured at $1.4\times10^{-10}\,\text{g}^{-1}$ (where g is the acceleration due to Earth's gravity). After machining we were able to minimize the sensitivity along the other degrees of freedom, without degrading the vertical sensitivity. Indeed, we initially measured a large sensitivity to accelerations in the longitudinal and transverse directions at about $2\times10^{-9}\,\text{g}^{-1}$. Before closing the vacuum chamber, we controllably shook the cavity along the different degrees of freedom using the modulation port of a commercial active vibration isolation (AVI) system, with the laser locked to the cavity and compared to another reference ultrastable laser. We employed a modulation frequency of 8~Hz, at which we could selectively excite each degree of freedom and impart an acceleration amplitude of $\simeq5\times10^{-3}\,\text{g}$. As these measurements were not taken in vacuum, we found that closing the cavity's venting holes with Kapton tape was crucial to minimize pressure-induced frequency noise. We were able to reduce the sensitivity to transverse acceleration to $3\times10^{-10}\,\text{g}^{-1}$ by applying transverse forces to the self-balancing structure~\cite{Hafner:15} of the cavity and to the cavity itself, with aluminum and Viton constraints, respectively, as shown in Fig.~\ref{fig:1}(b) and \ref{fig:1}(d). We measured that by placing tuning masses directly on the top surface of the cavity, we could tune the sensitivity to longitudinal acceleration to $2\times10^{-10}\,\text{g}^{-1}$, as displayed in Fig.~\ref{fig:1}(e). We expect that these constraints and tuning masses are fine tuning the distribution of weight on the four Viton disks supporting the cavity, shown in Fig.~\ref{fig:1}(a), symmetrizing the reaction forces and compensating for any residual asymmetry introduced in the machining or positioning of the cavity. After this tuning, we closed the chamber and started the vacuum pumps. We observed that while the vertical and transverse sensitivities remained constant over time, the longitudinal sensitivity started to increase, finally stabilizing at a value of $8\times10^{-10}\,\text{g}^{-1}$ after about six months, possibly due to deformation of the supporting Viton disks and/or self-balancing structure. The vacuum chamber hosting the cavity is secured on a honeycomb optical breadboard, which sits on a commercial AVI system. We achieved the best vibration isolation above 0.1~s averaging time by combining the outputs of the commercial sensors of the AVI with the velocity outputs of a seismometer and angle variations measured by a tiltmeter, both placed on the breadboard, as shown in Fig.~\ref{fig:1}(c) and \ref{fig:1}(f). More details of ULE68a's design and custom AVI are in the Supplement.
The vacuum is maintained by five ion pumps and five non-evaporable getters, with pumping speeds of about 50~liters s$^{-1}$ for nitrogen and 400~liters s$^{-1}$ for hydrogen, respectively. This redundant approach is chosen to operate the pumps far away from saturation, and to be able to switch off one pump in case of unstable behavior. A pressure of $2\times10^{-8}\,$mbar is achieved, measured at the ion pumps and with a contribution to frequency instability below the thermal noise floor. The vacuum chamber contains three low-emissivity polished aluminum thermal shields to increase the passive thermal insulation of the cavity, resulting in a characteristic thermal response time of about ten days. Heaters attached on the exterior surface of the chamber stabilize the temperature of the system at 26$\,^{\circ}$C, a few degrees above room temperature, with mK temperature instability. ULE68a operates at 1542~nm, to leverage on the availability of high quality components (optics, lasers, optical amplifiers) at telecom wavelengths, with the additional benefits of producing high signal-to-noise-ratio optical beats with frequency combs and simplifying distribution of ultrastable light through optical fibers. The 1542~nm laser is stabilized to a TEM$_{00}$ mode of ULE68a with a bandwidth of 290~kHz, and with full path length stabilization using the cavity transmission as reference. To selectively measure the stability of ULE68a we realized a three-cornered hat (TCH) scheme~\cite{Allan:74} with ULE48a and ULE48b, as described in Fig.~\ref{fig:2}(a). Both ULE48a and ULE48b work at 1064~nm, therefore a frequency comb is employed to bridge the wavelength gap with ULE68a at 1542~nm. ULE48a, ULE48b and the frequency comb are located in a different building, about 1~km distant from the building hosting ULE68a. The comb is stabilized to ULE48a and its component at 1542~nm is transferred to ULE68a's location through a 1~km long optical fiber, generating the beat \text{"ULE68a–ULE48a"} at 1542~nm, captured on a frequency counter. The beat \text{"ULE48a–ULE48b"} at 1064~nm is measured on a different frequency counter. The counters are synchronized within 100~ms, have zero dead time and are operated with 0.5~s integration time and in $\Lambda$ counting mode. The third beat frequency \text{"ULE68a–ULE48b"} is calculated from the difference of the other two beats, which is justified since the noise of our beat measurement system is negligible. For this calculation all the beats are expressed in fractional units. We measure that ULE48a, ULE48b and ULE68a reach a fractional frequency instability of $7\times10^{-17}$, $5.5\times10^{-17}$ and $4\times10^{-17}$, respectively, between 5~s and 20~s averaging time, as shown in Fig.~fig:2(b). To the best of our knowledge, ULE68a's instability is the lowest ever reported for an ultrastable laser based on an optical reference cavity at room temperature~\cite{Ludlow:15, Young:99, Jiang:11, Hafner:15, Schioppo:17, Schioppo:22}, and it is competitive with the best stability demonstrated by cryogenic realizations~\cite{Wiens:14, Matei:17, Kedar:23, Valencia:24, Chen:25, Lee:2026}. At 0.5~s, the instability of ULE68a at $1.2\times10^{-16}$ approaches known noise at $1.1\times10^{-16}$, which accounts for vibrations, thermal noise, and instability transfer due to the comb and fiber link. The instability reported for our single-branch comb is not a fundamental limit, and it can be improved by minimizing the amount of non-common-mode optical path-length ﬂuctuations occurring in the beats with the comb lines. Additional measured instability could be attributed to counting noise arising from synchronization and aliasing effects~\cite{Benkler:19}. Instability above 100~s is limited by non-linear length variation of the cavities. In this implementation we did not use compensating ULE-rings~\cite{Legero:10} for ULE68a, to minimize deviations from the FEM acceleration sensitivity model introduced by imperfect alignment of spacer, mirrors and rings in the optical contacting. Due to the long thermal response time we did not characterize the zero-CTE temperature and we simply stabilized ULE68a at a few degrees above room temperature. Since we have demonstrated that tuning the sensitivity to accelerations is possible with constraints and masses, the addition of ULE compensating rings in future designs could improve stability above 100~s averaging time. The excess noise that limits instability at $4\times10^{-17}$, above the expected thermal noise of $2.5\times10^{-17}$, could be associated with local deformations at the spacer ends, near the mirrors, due to the absence of compensating ULE rings \cite{KesslerTherm:12}. Additional thermal noise could arise from rubber-type material with low mechanical quality factor~\cite{KesslerTherm:12} used in the self-balancing structure and transverse constraints.
\begin{figure*}[htb]
\centering
\includegraphics[width=0.94\linewidth]{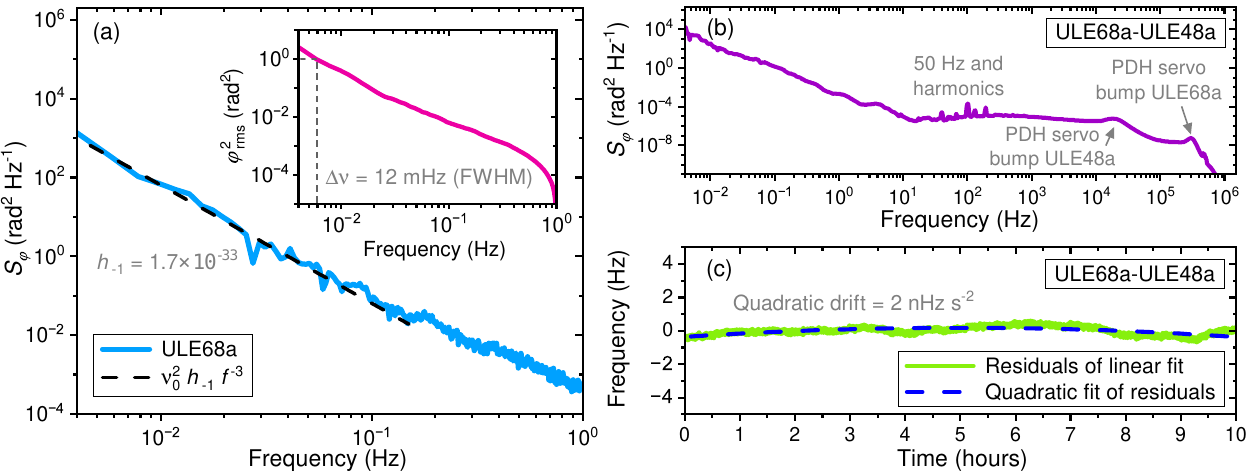}
\caption{(a) Phase noise power spectral density $S_{\varphi}$ of ULE68a, extracted with the three-cornered hat method, starting from the time series of the beats. The black dashed line represents the flicker frequency noise $\nu^2_0\,h_{-1}\,f^{-3}$, with $h_{-1}=1.7\times10^{-33}$ and $\nu_0$ the optical carrier frequency for the wavelength 1542~nm. The inset plot shows the rms phase noise integration of $S_{\varphi}$. Gray short-dashed lines indicate where $\varphi^{2}_{\text{rms}}=\,1\,\text{rad}^{2}$, achieved after an integration down to 6~mHz, corresponding to a laser linewidth full width at half maximum of 12~mHz. (b) Phase noise measurement of ULE68a-ULE48a beat, taken with a phase noise analyzer, linear drift removed. (c) Residuals of the linear fit of the frequency beat ULE68a-ULE48a (green line), with quadratic fit (blue dashed line).}
\label{fig:3}
\end{figure*}
Starting from the time series of the beats, the TCH method also allows us to extract the phase noise power spectral density $S_{\varphi}$ of each individual laser involved in the scheme (method in Supplement). In Fig.~\ref{fig:3}(a) we report $S_{\varphi}$ of ULE68a, where the high frequency cut-off at 1~Hz is given by the counting interval set at 0.5~seconds. The integration of the measured phase noise can be used to provide an estimation of the laser linewidth full width half maximum $\Delta\nu$ according to $\int_{\Delta\nu/2}^{+\infty} S_{\varphi} \,df= 1\,\text{rad}^2$~\cite{Walls:75}, which gives a value $\Delta\nu=12\,$mHz, as displayed in the inset of Fig.~\ref{fig:3}(a), at the same level as state-of-the-art cryogenic cavities~\cite{Matei:17, Chen:25}. Contributions to laser linewidth in the integration of $S_{\varphi}$ become more relevant at frequencies below 100~mHz, so starting from this value we estimate the $h_{-1}$ parameter accounting for the flicker frequency noise that describes the thermal noise limit, as shown in Fig.~\ref{fig:3}(a). For ULE68a, we find $h_{-1}=1.7\times10^{-33}$ which is consistent with the TCH estimation of the flicker floor at $4\times10^{-17}$, through $\sigma_y\simeq\sqrt{0.935\,h_{-1}}$. In Fig.~\ref{fig:3}(b) we show $S_{\varphi}$ of the ULE68a-ULE48a beat. Operation at room temperature ensures an uneventful spectrum, without any major frequency components except the PDH frequency stabilization servo bumps. The reported $S_{\varphi}$ measurements also include the contributions of the comb and fiber link. Residuals of the linear fit of the beat ULE68a-ULE48a during the TCH are shown in Fig.~\ref{fig:3}(c), modeled as a quadratic behavior with a coefficient of 2~nHz$\,\text{s}^{-2}$, accounting for both cavities. Linear drift of ULE68a is isolated to 19~mHz$\,\text{s}^{-1}$ (see Supplement), typical of  state-of-the-art room temperature ULE cavities~\cite{Hafner:15,Schioppo:17, Schioppo:22}. \\ \indent
In conclusion, we have demonstrated the lowest fractional frequency instability ever reported for an ultrastable laser referenced to a room temperature optical cavity, and competitive with state-of-the-art cryogenic realizations. We discussed present limitations and possible routes for improvement. This work shows that state-of-the-art laser frequency stability and spectral purity are achievable at  room temperature, supporting and enabling applications that require continuous operation and reduced maintenance time, personnel and costs, making this optical frequency stability performance available to a wider community. 
\begin{backmatter}
\bmsection{Funding} UK Government Department for Science, Innovation and Technology through UK National Quantum Technologies Programme.
\bmsection{Disclosures} The authors declare no conflicts of interest.
\bmsection{Data availability} Data underlying the results presented in this paper may be obtained from the authors upon reasonable request.
\bmsection{Supplemental document}
See Supplement for supporting content.
\end{backmatter} 
\bibliography{bibliography}

{\huge\textbf{~~~~~~~~~~~~~~~~~~~Supplement}}

\section{Cavity design and experimental set-up}

\subsection{Design to minimize sensitivity to accelerations}

\begin{figure*}[bh!]
\centering
\includegraphics[width=0.99\linewidth]{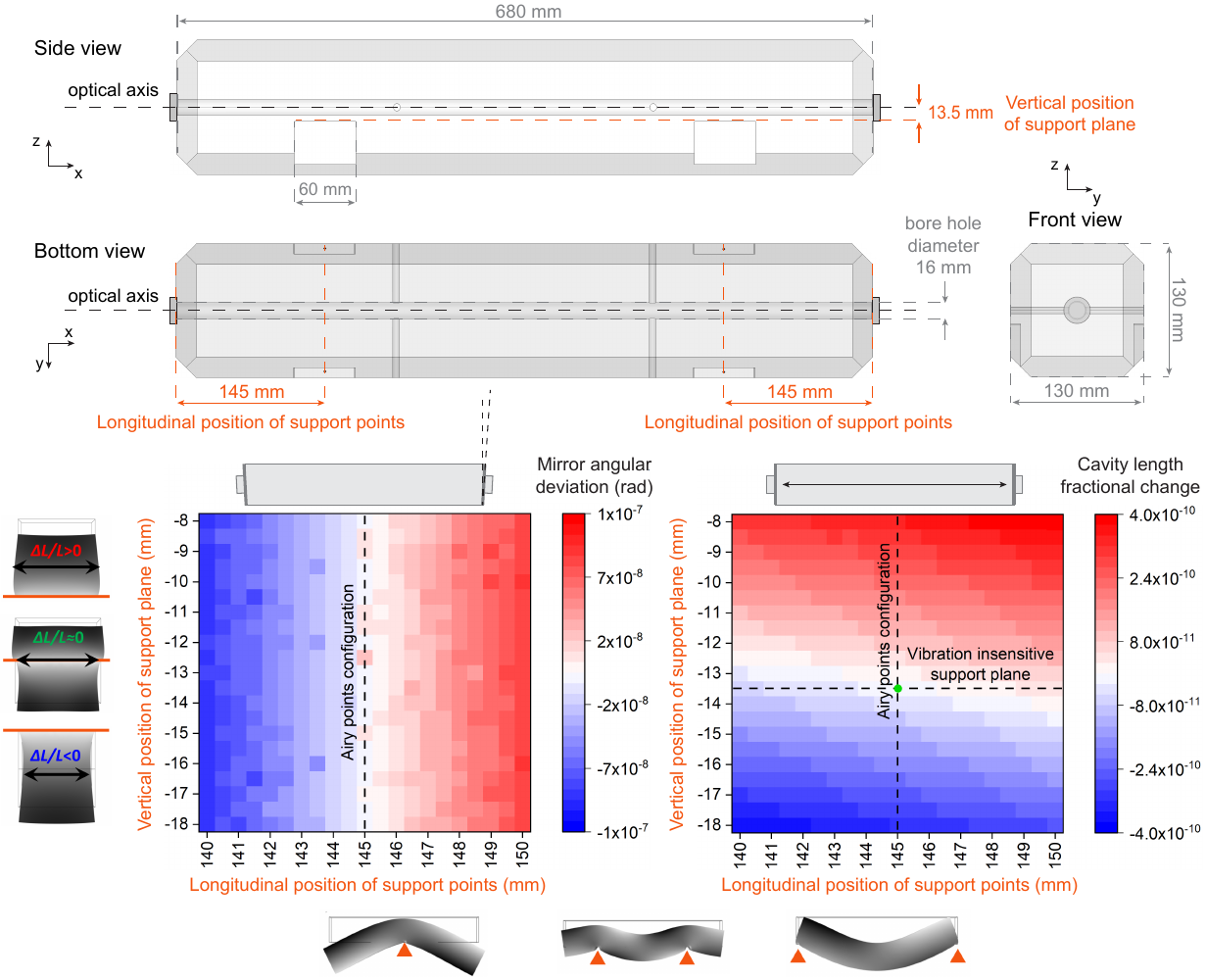}
\caption{Drawings and finite-element-method (FEM) simulations performed to minimize the sensitivity to accelerations of the ULE68a cavity.}
\label{Fig.S1}
\end{figure*}

To reduce sensitivity to accelerations, the geometry of the cavity is carefully optimized based on finite-element-method (FEM) simulations in the design phase, whilst also considering practical limitations in the machining phase. As explained in the main text, the chosen cuboid geometry gives us better consistency with the FEM simulations, because of reduced defects during the machining process. FEM simulations are carried out as a function of two main parameters critically impacting the sensitivity of the cavity to accelerations: the longitudinal position of the support points and the vertical position of the support plane, graphically described in the cavity drawings and FEM analysis of this 2D-parameter space in Fig.~\ref{Fig.S1}.

With FEM analysis, we first find the ideal longitudinal position of the supporting points, generally referred to as the "Airy points" configuration (in mechanical engineering), where the angular tilts of the two end faces of the spacer (and therefore of the mirrors), from deformation under its own weight, are null. Operating at the Airy points configuration is important as in practice we always expect a finite mismatch between the geometrical and the optical axis of the cavity, caused by finite tolerances in parallelism, which ends up converting tilts into effective distance variation between the points where the light beam touches the mirrors. In simpler words, the optical standing wave inside the cavity is, in reality, centered on the mirrors with a finite offset that depends on the parallelism of the cavity, and this coverts mirror tilts into optical path variations. 
\begin{figure*}[t!]
\centering
\includegraphics[width=0.99\linewidth]{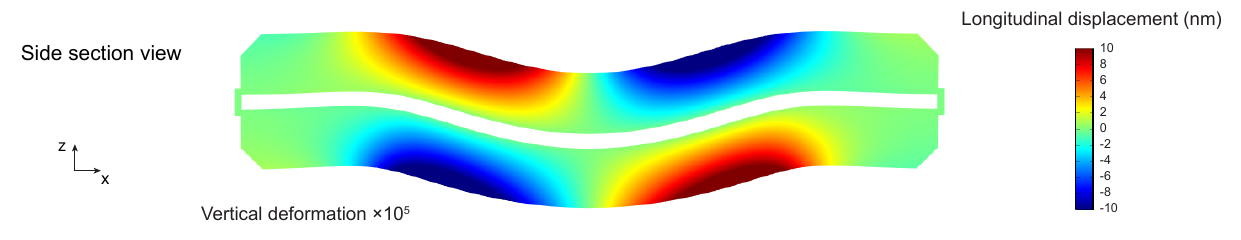}
\caption{Side section view of FEM analysis for an optical cavity geometry of ULE68a optimized for minimal sensitivity to accelerations. The vertical deformation is shown exaggerated by $10^5$, while the longitudinal displacement is represented by the color scale on the right of the figure. Both the Airy points configuration (null angular tilt of the cavity ends) and anti-symmetric longitudinal displacement are achieved.}
\label{Fig.S2}
\end{figure*}

Once the Airy points configuration is found, we run FEM analysis to look at the change of the cavity length. In this last step we seek the vertical position of the support plane that produces equal and opposite longitudinal displacements of different parts of the spacer, canceling out at the mirrors location. The optimal design of the cavity is realized by these two conditions being achieved simultaneously, namely, the Airy points configuration and the anti-symmetric longitudinal displacement. A detailed view of this optimized configuration is provided in Fig.~\ref{Fig.S2}, where we display a section view of the cavity, with vertical deformation exaggerated by $10^5$ to show the good parallelism of the end faces of the cavity, and with the colors conveying the information about the longitudinal (along the optical axis) displacement, which is anti-symmetric in proximity of the areas where the cavity is supported and null at the cavity mirrors.
\begin{figure*}[b!]
\centering
\includegraphics[width=0.45\linewidth]{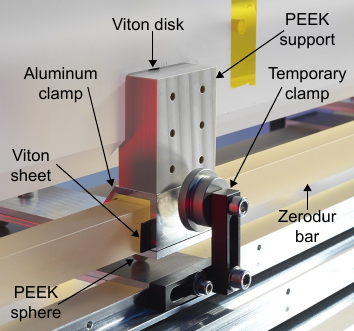}
\caption{Photo of one of the four pillars supporting the ULE68a cavity. A Viton disk mediates contact between the cavity spacer and the PEEK supporting pillar. An aluminum clamp holds the PEEK pillar to the two Zerodur bars that run alongside the cavity. Below this, there is a PEEK sphere between the aluminum clamp and the base of the innermost thermal shield. The extra arrangement of post and clamps seen on the side of the PEEK support was a temporary measure to keep the cavity secure before the self-balancing structure was finalized.}
\label{Fig.S3}
\end{figure*}

To support the cavity spacer on the chosen plane, four rectangular cut-outs are removed from the sides of the spacer, as seen in Fig.~\ref{Fig.S1}. The cavity sits on Viton disks to mediate contact between the cavity cut-outs and PEEK support pillars, with this detail shown in Fig.~\ref{Fig.S3}. The rectangular cut-outs and PEEK supports are sufficiently large to allow for an easy positioning of the Viton disks and for an experimental optimization of their position aimed to reduce the sensitivity to accelerations.
The Viton disks used have a diameter of 4.5~mm and a thickness of 1.6~mm. Disks were chosen over more traditional spheres due to better repeatability of their positioning and compression, allowing for more reliable experimental optimization of the supporting points to minimize sensitivity to accelerations. In contrast, attempts with Viton spheres proved problematic, as rolling of the spheres during positioning made it challenging to place all four in their desired positions repeatably. Viton hemispheres were also investigated, but with similar limitations in terms of reproducibility of positioning and compression. As described in the main text, in the final configuration the minimization of the sensitivity to accelerations was achieved with tuning masses and lateral constraints as overall this approach proved to be more effective and reproducible than moving the Viton disks. 

The anti-symmetric displacement configuration described in Fig.~\ref{Fig.S2} is realized in practice when the reaction forces on the supporting points are equal. Since four points over-define a plane, a state-of-the-art horizontal cavity needs to sit on a self-balancing support structure, which provides the needed compliance to symmetrize these reaction forces and to make sure that the four Viton disks are compressed by the same amount, as originally described and implemented in Ref. \cite{Hafner:15}. Our self-balancing structure is indeed inspired by this work. Two Zerodur bars are used to set the longitudinal distance between the aluminum clamps connected to the PEEK pillars. In our implementation, we use Viton spheres and sheets as intermediate material between the Zerodur bars and aluminum clamps to provide additional compliance, distribute the forces and prevent damaging/cracking of the Zerodur bars. 

\subsection{Cavity spacer machining and length limitations}

\begin{figure*}[b!]
\centering
\includegraphics[width=0.9\linewidth]{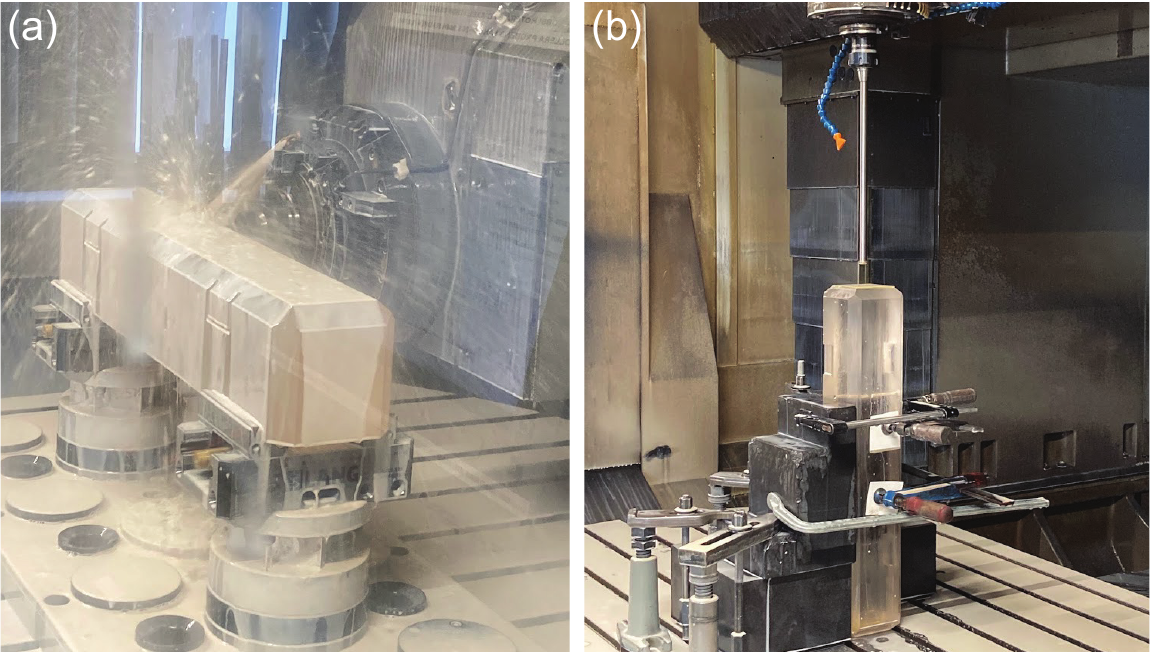}
\caption{Photos of ULE68a's cavity spacer during the machining process. (a) Milling of cuboid geometry. The cavity spacer is held securely and the milling tool moves in order to cut it. (b) Drilling of the bore hole for the optical path.}
\label{Fig.S4}
\end{figure*}

The cuboid spacer geometry used here was chosen based on feedback from the company that machined the three cavity spacers presented in this work, since this shape is less prone to defects than the original cylindrical geometry, as the cuboid geometry can be realized with a milling process, as shown in Fig.~\ref{Fig.S4} (a), and discussed in the main text. The feedback on practical limitations of the machining process was considered in the design phase of ULE68a to maximize the fidelity of the final machined spacer to the FEM model.

An additional machining consideration for long cavities is the drilling of the bore hole along the optical axis of the spacer. The alignment of this hole has an impact on the symmetry of the cavity, and therefore its sensitivity to accelerations. To ensure good alignment, the cavity is held vertically during the drilling process, as shown in Fig.~\ref{Fig.S4} (b), so that the drill is not pulled away from the desired path by gravity, as would happen with horizontal drilling. Additionally, the cavity spacer is bored from both ends. If the hole were bored from only one side, the drill would be more likely to drift due to the greater distance and thus create asymmetry. The diameter of the bore hole must also be chosen carefully. A compromise between the better rigidity (and length availability) of a thicker drill bit whilst still having enough surface area for optical contacting of the mirrors led us to a bore hole diameter of 16~mm. 

\begin{figure*}[ht]
\centering
\includegraphics[width=0.9\linewidth]{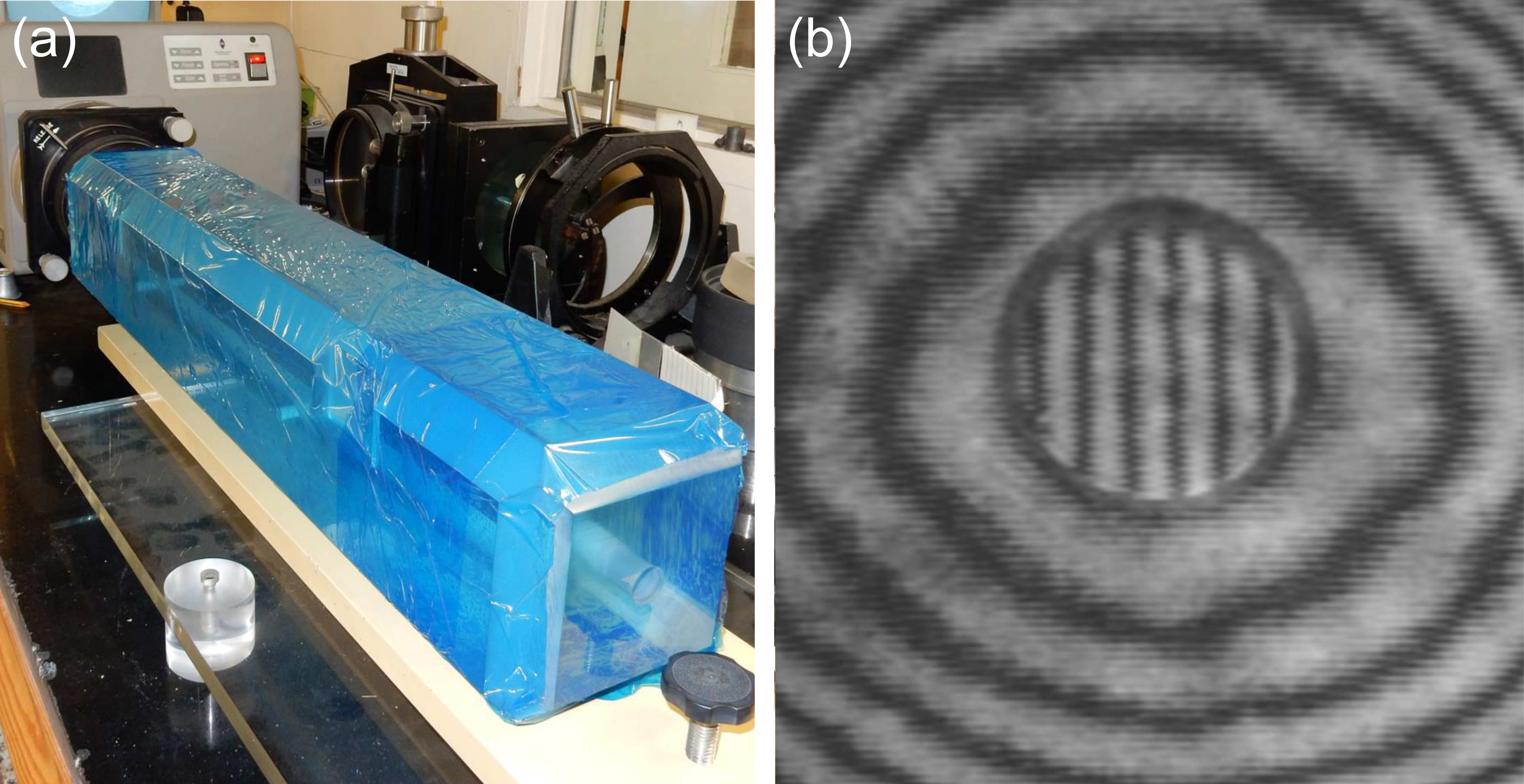}
\caption{(a) Photo of ULE68a during interferometric characterization of the flatness and parallelism of the cavity's end faces.  (b) Fringes on the area of the bore hole from the parallelism test. Counting of the fringes allows for calculation of the angle between the end faces.}
\label{Fig.S5}
\end{figure*}

After polishing, the cavity geometry is validated through interferometric evaluation of the flatness and parallelism of the end faces, as shown in Fig.~\ref{Fig.S5}. The deviation from flatness must be less than $\lambda /10$ to allow for proper optical contacting of the mirrors. For our 68~cm spacer with mirrors of a 10.2~m radius of curvature, we require deviation from parallelism below 1~arcmin ($\leq2.9\times10^{-4}$~rad). The level of parallelism achieved for this cavity was 0.8~arcmin (2.4$\times10^{-4}$~rad), corresponding to a beam displacement of 1.2~mm with respect to the center of the mirror, given the radius of curvature of the mirrors. This level of displacement is similar to possible offsets introduced in optical contacting the mirrors ($\simeq$ 0.5~mm for each mirror), and to the 1.9~mm diameter of the beam on the mirrors. The realized parallelism was estimated with an interferometric technique, counting 6 interference fringes (at the wavelength of the interferometer laser 633~nm) on the area on the bore hole, between the two end faces of the spacer, as shown in Fig.~\ref{Fig.S5}(b). 

Error in the parallelism also presents the first machining limitation on the extension of the cavity length. Increasing the length of the spacer further would require higher parallelism while also making it more challenging to achieve in machining. There are a few further practical limitations in increasing the length of a cavity spacer beyond the 68~cm chosen for this system. Generally, increasing the size of the spacer will make all tolerances harder to meet, increasing the machining costs and limiting the choice in manufacturers able to realize these requirements. Another factor that places a firm limit is the availability of ULE blanks long enough to create a monolithic cavity, keeping in mind that the machining and polishing processes both remove small amounts of material. Due to the manufacturing methods of the ULE glass, blanks longer than around 1~m are not readily available. 
In terms of performance, the main limitation on cavity length is the escalating sensitivity to accelerations, which becomes more and more challenging to counteract in order to reach the lower thermal noise limit the longer spacer aims to provide.

\subsection{Custom vibration isolation system}
\begin{figure*}[ht]
\centering
\includegraphics[width=0.9\linewidth]{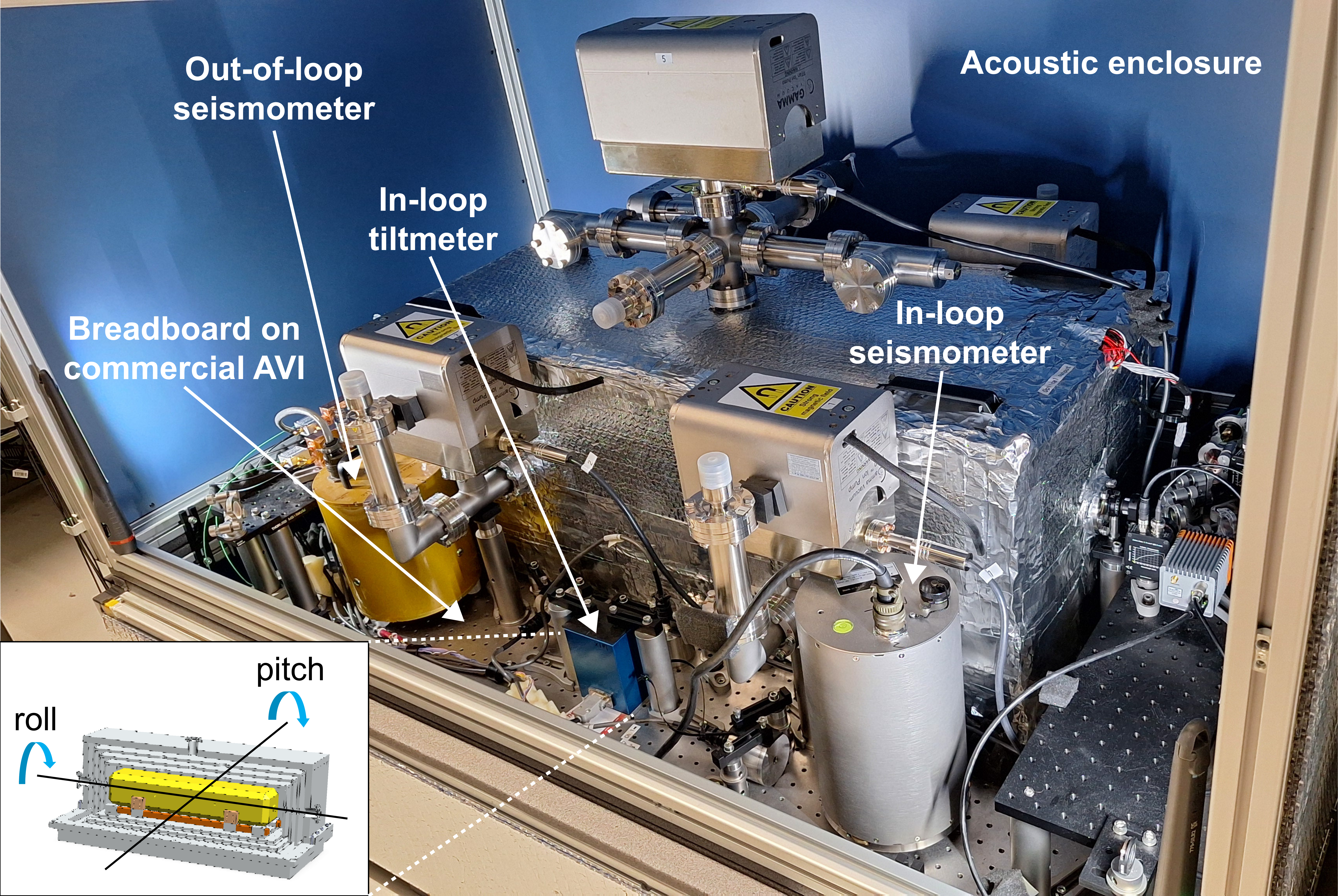}
\caption{Photo of the ULE68a set-up inside its acoustic enclosure. The photo was taken with one side of the acoustic enclosure open for access. The cavity is within the vacuum system on top of the optical breadboard, which sits on a commercial active vibration isolation (AVI) system. Also on the breadboard are two seismometers and one tiltmeter. The acceleration outputs of one seismometer are used for out-of-loop diagnostics, and the velocity outputs of the other contribute to our custom AVI set-up. The tiltmeter has two outputs, pitch and roll, which are also used in the custom AVI. The inset diagram clarifies the significance of these tilts in regard to the cavity geometry.}
\label{Fig.S6}
\end{figure*}
In Fig.~1(c) and (f) of the main text, we compare the acceleration noise experienced by the cavity in three different active vibration isolation (AVI) configurations: "no AVI", "commercial AVI", and "custom AVI". The "no AVI" configuration is purely passive. The "commercial AVI" configuration consists of a commercial AVI system set to its default gain, while the "custom AVI" is a combination of the commercial AVI, its low frequency sensor (LFS) extension, and custom feedback from in-loop sensors shown in Fig.~\ref{Fig.S6}. For the "custom AVI" implementation, the commercial systems are set to lower gains to avoid an increase of acceleration noise above 0.5~s averaging time. Specifically, the commercial AVI and LFS are set to $10\,\%$ and $25\,\%$ of their default gains, respectively. 
The custom feedback encompasses velocities and angle variations captured by a seismometer and a tiltmeter, respectively, both placed on the optical breadboard on which the cavity's vacuum chamber sits, as seen in Fig~\ref{Fig.S6}. These signals are filtered, scaled and summed together before being passed to the modulation ports of the commercial AVI system, which combines them with the information from the sensors of the commercial AVI and its LFS extension. With this ensemble of sensors, we improve vibration isolation above 0.1~s averaging time with respect to both the "no AVI" and "commercial AVI" configurations. 
The acceleration spectrum and modified Allan deviation of acceleration time series provided in the main text in Fig. 1(c) and Fig. 1(f) are for the longitudinal direction where we have the maximum sensitivity. The accelerations along the transverse and vertical directions have very similar behavior to the longitudinal degree of freedom. 

\section{Three-cornered hat method}

The three-cornered hat method is a well established technique for separating individual instability contributions from beat measurements of three independent oscillators ~\cite{Allan:74}. For frequency oscillators $a$, $b$, and $c$ with similar noise profiles, starting from the beats measured for the pairs $ab$, $ac$, and $bc$, we can derive the related Allan variances
$\sigma_{y, ab}^2\,$, $\sigma_{y, ac}^2$ and $\sigma_{y, bc}^2$. In the limit of independent oscillators the following decoupling equations can be used to extract the Allan variance $\sigma_{y, x}^{2}$ of the individual oscillator $x$
\begin{equation*}
\begin{aligned}
\sigma_{y, a}^{2} &= \frac{1}{2}(\sigma_{y, ab}^{2}+\sigma_{y, ac}^{2}-\sigma_{y, bc}^{2}) \\
\sigma_{y, b}^{2} &= \frac{1}{2}(\sigma_{y, ab}^{2}+\sigma_{y, bc}^{2}-\sigma_{y, ac}^{2}) \\
\sigma_{y, c}^{2} &= \frac{1}{2}(\sigma_{y, ac}^{2}+\sigma_{y, bc}^{2}-\sigma_{y, ab}^{2})\,.
\end{aligned}
\label{eq:S1}
\end{equation*}
 The Allan deviation $\sigma_{y, x}$ is simply the square root of the Allan variance. These relations hold without loss of generality for the Modified Allan deviation (MDEV), as used in the measurement presented in Fig.~2 (b) of the main text. 

Furthermore, this treatment can be extended to fractional frequency noise power spectral density $S_{y}$, since, like $\sigma_{y}^{2}$, it scales linearly with the $h_n$ parameters used to capture different types of noise, according to $S_{y}(f)=\sum_{n}h_n f^n$, where $f$ is the Fourier frequency.   
In this work, experimental $S_{y,ab}\,$, $\,S_{y,ac}$ and $S_{y,bc}$ were extracted from the frequency beats pairs (time series measurements) $ab$, $ac$, and $bc$, with $a$, $b$, and $c$ representing the cavities ULE68a, ULE48a and ULE48b, respectively. The following decoupling equations were used to determine the $S_{y}$ of the individual cavities 
\begin{equation*}
\begin{aligned}
S_{y, a} &= \frac{1}{2}(S_{y, ab}+S_{y, ac}-S_{y, bc}) \\
S_{y, b} &= \frac{1}{2}(S_{y, ab}+S_{y, bc}-S_{y, ac}) \\
S_{y, c} &= \frac{1}{2}(S_{y, ac}+S_{y, bc}-S_{y, ab})\,.
\end{aligned}
\label{eq:S2}
\end{equation*}
For Fig.~3 (a) and the subsequent analysis in the main text, the fractional frequency noise power spectral density (PSD) of ULE68a has been converted to phase noise PSD $S_{\varphi}$, using the expression
$S_{\varphi}(f)=(\nu_{0}/f)^2S_{y}(f)\,$,
where $\nu_{0}$ is the optical carrier frequency.

\section{Linear drift}
\begin{figure*}[ht!]
\centering
\includegraphics[width=0.9\linewidth]{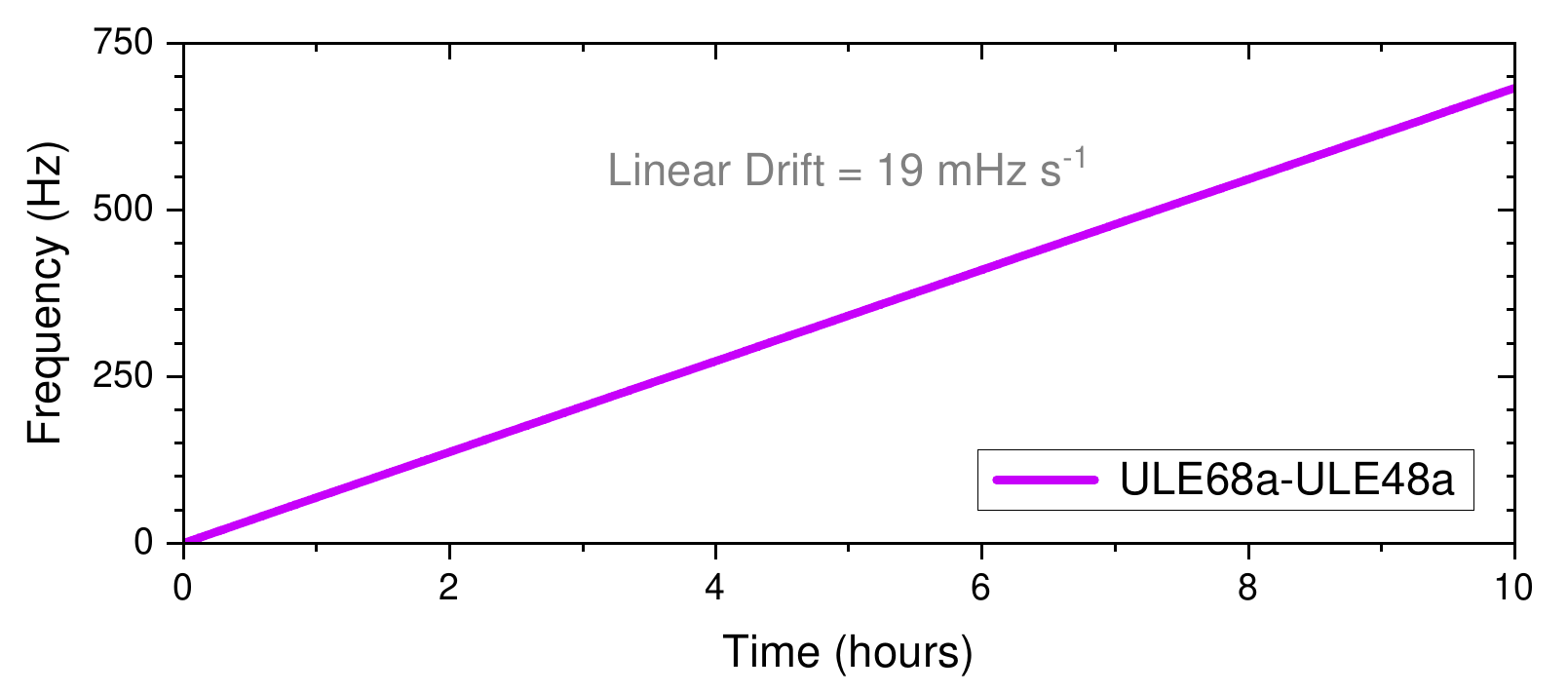}
\caption{Frequency beat ULE68a-ULE48a during the three-cornered hat measurement, with offset removed and counter integration time of 0.5~s. The linear drift of ULE48a is hardware-removed at the beginning of this measurement.}
\label{Fig.S7}
\end{figure*}

The three-cornered hat measurement presented in Fig.~2 of the main text was taken over a continuous period of 10~hours. The linear drift of ULE48a is hardware-removed at the mHz s$^{-1}$ level, measured with an optical atomic clock. The hardware de-drift rate for ULE48a is set at the beginning of the measurement of the beat ULE68a-ULE48a. In this way we isolate the linear drift of ULE68a in the counted beat ULE68a-ULE48a, to a value of $\simeq$19~mHz$\,\text{s}^{-1}$, as seen in Fig.~\ref{Fig.S7}, with a tolerance of a few mHz s$^{-1}$. The sign of the drift is consistent with the cavity shrinking over time, as expected for a glass material. The residuals after linear drift removal are shown in Fig.~3(c) of the main text, where we estimate a combined ULE68a-ULE48a non-linear drift component of 2~nHz$\,\text{s}^{-2}$.

\end{document}